\documentclass[a4paper,twoside,twocolumn]{article}
\usepackage{multicol,multirow,graphics,fancyhdr,epstopdf}
\usepackage{amsmath,amsfonts,amssymb,bm,upgreek,mathrsfs,ccmap,mathcomp}
\usepackage[pagewise,switch,columnwise]{lineno}
\usepackage[compress,nospace]{cite}
\usepackage[dvipsnames]{xcolor}
\usepackage{CPL-2022}
\usepackage{ccmap}
\usepackage{braket}
\usepackage{graphicx}
\usepackage{multicol}
\usepackage[font=small,skip=5pt]{caption}
\usepackage{hyperref}
\usepackage{CJKutf8}

\begin{document}

\twocolumn[\begin{@twocolumnfalse}
\begin{center}
\large\bf{\boldmath{
$P_c(4457)$ interpreted as a $J^P=1/2^+$ state by $\bar{D}^0\Lambda^+_c(2595) - \pi^0 P_c(4312)$ interaction 
}}\\
\footnotetext{\hspace*{-5.4mm}$^{*}$Corresponding authors. Email:  jypang@usst.edu.cn; wujiajun@ucas.ac.cn

\href{http://www.iop.org}{IOP Publishing Ltd}}
\normalsize \rm{}
\begin{CJK*}{UTF8}{gbsn}

Jin-Zi Wu(吴金姿)$^{1,2}$, Jin-Yi Pang(庞锦毅)$^{3}$, and Jia-Jun Wu(吴佳俊)$^{1,4*}$\\
\end{CJK*}
\small{$^{1}$School of physics sciences, University of Chinese Academy of Sciences, Beijing 100049, China

$^{2}$Columbian College of Arts \& Sciences, George Washington University\\ 801 22nd St. NW Washington, DC 20052 

$^{3}$College of Science, University of Shanghai for Science and Technology, Shanghai 200093, China

$^{4}$Southern Center for Nuclear-Science Theory, Institute of Modern Physics, Chinese Academy of Sciences, Huizhou 516000, China}
\vskip 1.5mm

\small{\narrower 
$P_c(4457)$ has been discovered over five years, but the parity of this particle remains undetermined.
In this letter we propose a new interpretation for $P_c(4457)$, which is the state generated from the coupled-channel $\bar{D}^0\Lambda_c^{+}(2595)$ and $\pi^0 P_c(4312)$ since they can exchange an almost on-shell $\Sigma_c^+$.
In this scenario, the parity of $P_c(4457)$ will be positive, which is different from the candidate of the bound state of $\bar{D}^*\Sigma_c$.
The main decay channel of $P_c(4457)$ in this model is $P_c(4312)\pi$. 
We propose three processes $\Lambda_b^0 \to J/\psi K_s p \pi^-$, $\Lambda_b^0 \to J/\psi K^- p \pi^0$, and $\Lambda_b^0 \to J/\psi p \pi^- \pi^+ K^-$ to verify $P_c(4457)\to P_c(4312)\pi$.

\par}\vskip 3mm
\end{center}
\noindent{\narrower{DOI: \href{http://dx.doi.org/10.1088/0256-307X/41/9/091201}{10.1088/0256-307X/41/9/091201}}

\par}
\vskip 5mm
\end{@twocolumnfalse}]

\newcommand{\mev}{\textrm{ MeV}}
\newcommand{\gev}{\textrm{ GeV}}



In 2015, the LHCb Collaboration first reported two pentaquark-like resonances in the invariant mass spectrum of $J/\psi p$ in the reaction of $\Lambda_b^0 \to J/\psi K^- p$ decay~\cite{Aaij:2015tga,Aaij:2015fea}.
One of them was labeled as $P_c(4380)$ with a large width of about $205\mev$, while the other was labeled as $P_c(4450)$ with a small width of $39\mev$. 
These two states were also confirmed by a model-independent re-analysis of the experimental data~\cite{Aaij:2016phn}.
Furthermore, in another reaction of $\Lambda_b^0 \to J/\psi p \pi^-$ decay \cite{Aaij:2016ymb}, these two states were reconfirmed.
However, in 2019, the LHCb Collaboration obtained the Run-2 data, in which events of the reaction $\Lambda_b^0 \to J/\psi K^- p$ increased with a factor of 9~\cite{Aaij:2019vzc}. 
Three $P_c$ states were then clearly found~\cite{Aaij:2019vzc},
\begin{align*}
&M_{P_{c}(4312)} = (4311.9\pm0.7^{+6.8}_{-0.6})\mev,\\
&
  \Gamma_{P_{c}(4312)}= (9.8\pm2.7^{+3.7}_{-4.5})\mev,  \\
&M_{P_{c}(4440)} = (4440.3\pm1.3^{+4.1}_{-4.7})\mev,\\
&
  \Gamma_{P_{c}(4440)}= (20.6\pm4.9^{+8.7}_{-10.1})\mev, \\
&M_{P_{c}(4457)} = (4457.3\pm0.6^{+4.1}_{-1.7})\mev,\\
&
  \Gamma_{P_{c}(4457)}= (6.4\pm2.0^{+5.7}_{-1.9})\mev.
\end{align*}
Compared to the previous two $P_c$ states, $P_c(4450)$ was split into two states, $P_c(4440)$ and $P_c(4457)$, in addition to a new narrow resonance $P_c(4312)$. 
However, the updated results could neither confirm nor refute the broad earlier resonance $P_c(4380)$.
A theoretical work also suggested that $P_c(4380)$ may not be necessary for explaining the data published in 2015~\cite{Roca:2016tdh}.
Then in 2022, the LHCb Collaboration discovered a new state in the $J/\psi p$ invariant mass distribution with a mass of $4337^{+7-4}_{+2- 2}\mev$ and a width of $29^{+26-12}_{+14-14}\mev$ in another reaction $B_s^0\to J/\psi p\bar{p}$~\cite{LHCb:2021chn}, although its significance is in the range from $3.1$ to $3.7 \sigma$.
There are four $P_c$ states discovered by the LHCb collaboration until now. 
However, spin and parity have not been determined for any of these states.

On the theoretical side, various works explained these $P_c$ states.
Typically, before the first experimental discovery of the $P_c$ states in 2015,
there were several papers predicting these states, based on several different models~\cite{Wu:2010jy,Wu:2010vk,Wang:2011rga,Yang:2011wz,Yuan:2012wz,Wu:2012md,Garcia-Recio:2013gaa,Xiao:2013yca,Uchino:2015uha,Karliner:2015ina}.
After 2015, this topic aroused great interest among people.
The first experiment was cited more than 1700 times to the present time, a number of comprehensive articles have been published and here we list some of them~\cite{Chen:2016qju, Hosaka:2016pey, Chen:2016spr, Lebed:2016hpi, Esposito:2016noz, Guo:2017jvc, Ali:2017jda, Olsen:2017bmm, Karliner:2017qhf, Yuan:2018inv, Liu:2019zoy, Brambilla:2019esw, Guo:2019twa, Yao:2020bxx, Barabanov:2020jvn, Chen:2022asf, Huang:2023jec, Ortega:2016syt, Azizi:2016dhy}. 
Among all of these theoretical studies, the molecular picture is the favorite one, since all of these states are close to the $\bar{D}^{(*)}\Sigma_c^+$ threshold.
In this picture, $P_c(4312)$ is just below the threshold of $\bar{D}\Sigma_c$, while $P_c(4440)$ and $P_c(4457)$ are below the threshold of $\bar{D}^*\Sigma_c$, and all of them are predicted as the bound states hold by the $S$-wave attractive interaction. 
Thus, in the molecular picture, all three states should have negative parity.
However, this predicted property has not yet been confirmed experimentally. 
In contrast, in the early work~\cite{Aaij:2015tga}, two widely favored solutions of spins and parities show that these two states should have opposite parities!
Yet, in their new result in 2019~\cite{Aaij:2019vzc}, however, they did not give a further argument for the parity of these three states. 
Thus, whether these three $P_c$ states have positive or negative parity remains an open problem!

In the previous work~\cite{Geng:2017hxc}, the inclusion of $\bar{D}^0\Lambda^+_{c}(2595)$ channel was initially proposed. (For simplify, we use $\Lambda^+_{c1}$ instead of $\Lambda^+_{c}(2595)$ in this letter.) This channel contains two particles with $J^P=0^-$ and $1/2^-$, with the threshold around $4450 \mev$.
Then they may be bound by $S-$wave interaction, thereby one can interpret the $P_c(4450)$ as a positive parity state.
After $P_c(4457)$ was found, in Ref.~\cite{Burns:2019iih}, they used similar method to reproduced the $P_c(4457)$ as a $J^P=1/2^+$ state by considering the coupled-channel effect $\bar{D}^0\Lambda^+_{c1} - \bar{D}^*\Sigma_c$ with $\pi$-exchange.
In the later work Ref.~\cite{Yalikun:2021bfm}, however, the authors made a comprehensive analysis of the coupled channel system of $\bar{D}^0\Lambda^+_{c1}$ and $\bar{D}^*\Sigma_c$, they argue that "It is found that the role of the $\bar{D}^0\Lambda^+_{c1}$ channel in the descriptions of the $P_c(4440)$ and $P_c(4457)$ states is not significant with the one boson exchange (OBE) parameters constrained by other experimental sources."
Here, we propose a novel interpretation that includes not only $\bar{D}^0\Lambda^+_{c1}$ but also $\pi^0 P_c(4312)$, where we consider $P_c(4312)$ as the bound state of $\bar{D}\Sigma_c$ with $J^P=1/2^-$. 
These two-coupled-channel system may generate $P_c(4457)$ by $S$-wave interaction as a $J^P=1/2^+$ state.
It is clear that the thresholds of these two channels are very close to $4457$ MeV, and if a sufficiently strong $S$-wave attractive force exists, $P_c(4457)$ can be naturally explained as a bound state of the two-coupled-channel system, resulting in a positive parity.
{\color{black} It is worth to mention that usually the coupled channels which are close to the mass of hadron should play an important role to study the nature of such hadron for analiticity and unitarity~\cite{Guo:2017jvc,Yamaguchi:2013ty, Dong:2020hxe, Dai:2023kwv,Zhang:2024qkg,Song:2023pdq}. 
Although the $\bar{D}^*\Sigma_c$ channel is also very close to the $P_{c}(4457)$, however, it is $P$-wave contribution if we assume the parity of $P_{c}(4457)$ to be positive.
Thus, we will do not include $\bar{D}^*\Sigma_c$ channel here.}

\begin{figure}[h]
    \centering
	\includegraphics[scale = 0.42]{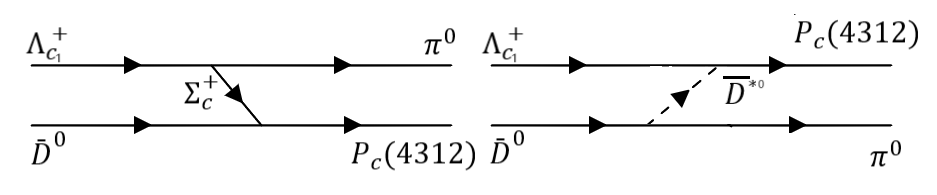}	
	\caption{The diagram of $\bar{D}^0\Lambda^+_{c1} \to \pi^0 P_c(4312)$ by exchanging $\Sigma_c^+$ and $\bar{D}^{*0}$.}
	\label{fig:exchange_diag}
\end{figure}

For the two-coupled-channel system of $\bar{D}^0\Lambda^+_{c1}$ and $\pi^0 P_c(4312)$, both of the diagonal interactions are neglected.
According to previous work~\cite{Yalikun:2021bfm}, the $\bar{D}^0\Lambda^+_{c1}\to \bar{D}^0\Lambda^+_{c1}$ process alone can not provide sufficient force to form a bound state, even when considering the contribution from the coupled-channel of $\bar{D}^{(*)}\Sigma_c$. 
For the interaction of $\pi^0 P_c(4312)\to \pi^0 P_c(4312)$, the leading order contact interaction Weinberg-Tomozawa term with $\pi$ is too weak to form a bound state because its proportionality to the very light mass of $\pi$.
On the contrary, the off-diagonal interaction between $\bar{D}\Lambda^+_{c1}$ and $\pi^0 P_c(4312)$ with $\Sigma_c^+$-state exchange, which is almost on-shell, may provide a sufficiently strong attractive interaction.
There are two possible mechanisms describing the off-diagonal interaction based on OBE, as shown in Fig.~\ref{fig:exchange_diag}.
Here we find the interaction by exchanging $\bar{D}^{*0}$ will be very small because of the small propagator of $\bar{D}^{*0}$ since this exchanged particle is deeply off-shell.
Therefore, in the following calculation, we only include the first diagram to verify the existence of a bound state in this system.

The potential between $\bar{D}^0\Lambda^+_{c1}$ and $\pi^0 P_c(4312)$ by $\Sigma_c^+$-exchange can be written as, 
\begin{equation}\begin{aligned}
     &\mathcal{V}_{\alpha\beta}(\boldsymbol{p},\,\boldsymbol{q},\, \lambda_{\alpha_B},\,\lambda_{\beta_B}) \\
     &= g_1 g_2\, \Bar{\mathcal{U}}_{\beta_B}(\boldsymbol{q},\lambda_{\beta_B}) 
     G^{\alpha\beta}_{\Sigma_c^+}(\boldsymbol{p},\boldsymbol{q})
     \mathcal{U}_{\alpha_B}(\boldsymbol{p},\lambda_{\alpha_B}).
\end{aligned}\end{equation}
The $\alpha$ and $\beta$ are the initial and final channel indexes, which could be the $\bar{D}^0\Lambda^+_{c1}$ and $\pi^0 P_c(4312)$.
The $\alpha_B$ and $\alpha_M$ are for the baryon and meson of the channel $\alpha$, respectively.
The three momenta $\boldsymbol{p}$ and $\boldsymbol{q}$ are the initial and final momenta of baryons, respectively.
Here $g_1$ and $g_2$ are the coupling constants for $\Lambda^+_{c1} \to \Sigma_c^+\pi^0$ and $P_c(4312) \to \bar{D}^0\Sigma_c^+$ respectively.
The wave function $\mathcal{U}_{i}$ is the baryon fields of the particle $i$ in the Dirac spinor representation defined as follows,
\begin{equation}\begin{aligned}
    \mathcal{U}_{i}(\boldsymbol{p}, \lambda_i) = \sqrt{\frac{\omega_i(\boldsymbol{p}) + m_{i}}{2m_{i}}}
    \begin{pmatrix}
    &\Phi^{\lambda_i}\\
    &\frac{\boldsymbol{\sigma}\cdot \boldsymbol{p}}{\omega_i(\boldsymbol{p})+m_{i}}\Phi^{\lambda_i}
\end{pmatrix},
\end{aligned}\end{equation}
where $\omega_i(\boldsymbol{p})=\sqrt{m_i^2+\boldsymbol{p}^2}$ is the on-shell energy with the mass of particle $i$, $m_i$; $\Phi^{\lambda_i}$ is the wave function of spin index $\lambda_i$.
The $G^{\alpha\beta}_{\Sigma_c^+}(\boldsymbol{p},\boldsymbol{q}, E)$ is the propagator of $\Sigma_c^+$, which is defined as follows,
\begin{equation}\begin{aligned}
     &G^{\alpha\beta}_{\Sigma_c^+}(\boldsymbol{p},\boldsymbol{q}, E) \\
     =&\frac{1}{2}\left\{ 
     \frac{ (\omega_{\alpha_B}(\boldsymbol{p})-\omega_{\beta_M}(\boldsymbol{q}))\gamma_0-(\boldsymbol{p}+\boldsymbol{q})\cdot\vec{\gamma} + m_{\Sigma_c^+}}{\left(\omega_{\beta_M}(\boldsymbol{q})-\omega_{\alpha_B}(\boldsymbol{p})\right)^2-\omega^2_{\Sigma_c^+}(\boldsymbol{p}+\boldsymbol{q})}
     \right.\\
     &\left.+
      \frac{ (\omega_{\beta_B}(\boldsymbol{q})-\omega_{\alpha_M}(\boldsymbol{p}))\gamma_0-(\boldsymbol{p}+\boldsymbol{q})\cdot\vec{\gamma} + m_{\Sigma_c^+}}
      {\left(\omega_{\beta_B}(\boldsymbol{q})-\omega_{\alpha_M}(\boldsymbol{p})\right)^2-\omega^2_{\Sigma_c^+}(\boldsymbol{p}+\boldsymbol{q})}
     \right\}.\label{eq:proSigmac}
\end{aligned}\end{equation}
Since our potential is designed to be energy independent, the propagator of $\Sigma_c^+$ will have two different forms as shown in the two terms in the brace of Eq.~(\ref{eq:proSigmac}), which is consistent with the prescription of Refs.~\cite{Wu:2012md}. 
Here we only consider the $S$-wave contribution of $V$, which is given by the following evaluation,
\begin{equation}\begin{aligned}
    &V_{\alpha\beta}\left(p,q\right)\\
    =&\frac{F(p,q)}{2\left(2\pi\right)^3}
\sqrt{\frac{m_{\Lambda^+_{c1}
}m_{P_{c}}}{\omega_{\Lambda^+_{c1}}(p) \omega_{P_c}(q) 2\omega_{D^0}(p) 2\omega_{\pi}(q)}} \\
&\cdot\ \sum_{\lambda_{\alpha_B},\,\lambda_{\beta_B}} 
2\pi\int_{-1}^1d \cos \theta d_{\lambda_{\alpha_B} \lambda_{\beta_B}}^{1/2}\left(\theta \right) 
\mathcal{V}_{\alpha\beta}(\boldsymbol{p},\,\boldsymbol{q},\,\lambda_{\alpha_B},\,\lambda_{\beta_B}).
\label{eq:valphabeta}
\end{aligned}\end{equation}
Here $F\left(p,q\right)$ is the form factor, 
\begin{equation}\begin{aligned}
    F\left(p,q\right) = \frac{\Lambda^2}{p^2 + \Lambda^2}\frac{\Lambda^2}{q^2+\Lambda^2},\label{eq:cut}
\end{aligned}\end{equation}
where $\Lambda$ is the energy cutoff.
The $d^{1/2}(\theta)$ is the Wigner-D matrix for spin-$\frac{1}{2}$ scattering. 


Then by the three-dimensional reduction of the Bethe-Salpeter equation (BSE), we can obtain the scattering equation for the partial wave scattering amplitude as follows, 
\begin{equation}\begin{aligned}
     &T_{\alpha \beta}\left(p,p';E\right)\\
     = & V_{\alpha\beta}\left(p,p'\right) + \sum_{\gamma} \int dq q^2 V_{\alpha\gamma}\left(p, q\right)G_{\gamma}\left(q; E\right)T_{\gamma \beta}\left(q, p';E\right),\label{equ:BSE_eq}
\end{aligned}\end{equation}
Only the off-diagonal elements of the potential $V-$matrix are nonzero, defined as Eq.~(\ref{eq:valphabeta}).
The meson-baryon propagator is 
\begin{equation}\begin{aligned}
    \mathcal{G}_{\gamma}(q;E) =\frac{1}{E-\omega_{\gamma_B}(q)-\omega_{\gamma_M}(q)+i\epsilon}.
\end{aligned}\end{equation}
Then by the standard numerical method, we can solve the matrix function of $T=(1-VG)^{-1}V$ to obtain the $T-$matrix based on the potential matrix $V$.
{\color{black} Here, $T$, $G$, and $V$ are the matrices of the momenta.
There is a standard numerical method to solve such scattering equation, please check Ref.~\cite{Wu:2012md}}.
Furthermore, through the analytical extension to the complex plane of the energy, the pole positions of the $T-$Matrix can be extracted from, 
\begin{equation}\begin{aligned}
   \det\left( \mathbb{I} - VG\right) = 0.
\end{aligned}\end{equation}
Such pole positions may correspond to the resonance, bound state or virtual state.

There are two coupling constants, $g_1$ and $g_2$, which were determined by the previous works.
The coupling constant $g_1$ was determined from the decay width of $\Lambda^+_{c1} \to \Sigma_c^+\pi^0$, with the notations above, $g_1$ can be calculated by the following expression,
\begin{equation}\begin{aligned}
     \Gamma &=  g_1^2 \frac{q_{on}}{4\pi}\frac{(\omega_{\Sigma_c^+}\left(q_{on}\right) + m_{\Sigma_c^+})}{m_{\Lambda^+_{c1}}}
\end{aligned}\end{equation}
Here the on-shell momentum $q_{on}$ satisfies $\omega_{\Sigma_c^+}(q_{on})+\omega_{\pi^0}(q_{on})=m_{\Lambda^+_{c1}}$ in the $\Lambda^+_{c1}$ rest frame, where we take $m_{\Lambda^+_{c1}}=2595\mev$, $m_{\Sigma_c^+}=2455\mev$ and $m_{\pi^0}=135 \mev$.
The world average of the decay width was determined to be $\Gamma = 2.2\mev$~\cite{Nakamura_2010}, thus the value of the coupling constant $g_1$ can be determined to be $g_1^2 = 0.4$.

The coupling constant of the $P_c\left(4312\right)$ and $\Sigma_c^+\Bar{D}^0$, $g_2$, was determined by the estimated $S$-wave coupling constant between the bound state and the corresponding coupled channel as follows,~\cite{Weinberg:1965zz, Baru:2003qq, Lin:2017mtz}
\begin{equation}\begin{aligned}
      g_2^2 = \frac{4\pi}{4m_{P_c}m_{\Sigma_c}} \frac{\left(m_{\Sigma_c}+m_{D}\right)^{\frac52}}{\left(m_{\Sigma_c}m_{D}\right)^\frac12}\sqrt{32\left|m_{\Sigma_c}+m_{D}-m_{P_c}\right|},
\end{aligned}\end{equation}
where we take $m_{P_c}=4312\mev$, $m_{\Sigma_c}=2455\mev$, and $m_{D}=1864.84\mev$.
Then we obtain $g_2^2=2.69$.

\begin{figure}[h]
    \centering
	\includegraphics[width=0.43\textwidth]{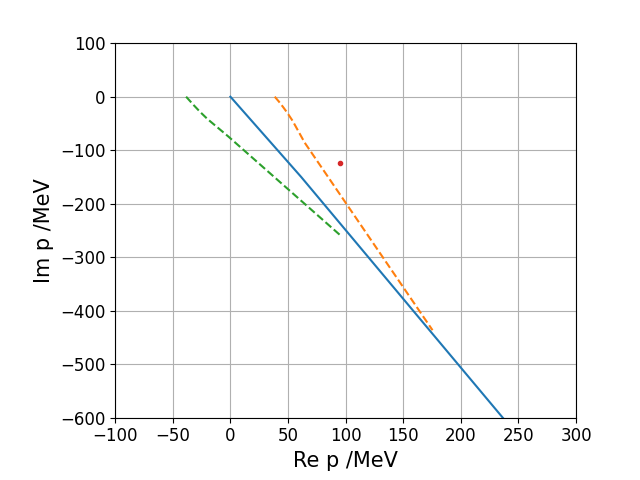}
        \includegraphics[width=0.43\textwidth]{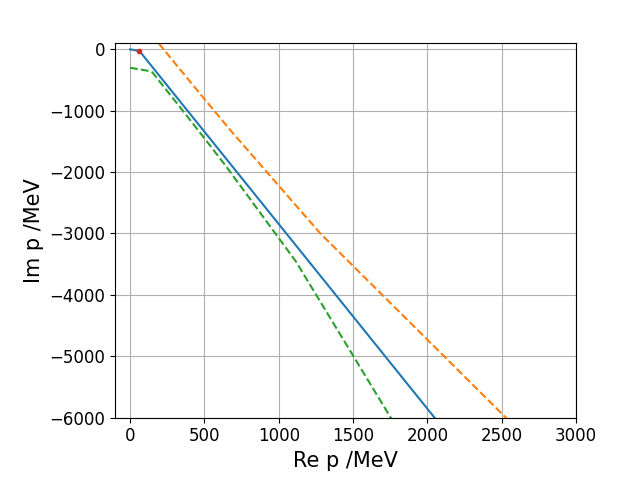}
      \caption{Two paths of integrate momenta for $\pi^0 P_c(4312)$(up) and $\bar{D}^0\Lambda^+_{c1}$(down).}
      	\label{fig:int_path}
\end{figure}

In this study, the main problem in the numerical calculation is the left-hand cut of the potential from the propagator of the exchange $\Sigma_c$ which will be very close to the on-shell.
Thus, if the value of $p$ is fixed in the function $V_{\alpha\beta}(p, q)$, there might exist $q=q_0$ which leads to the singular of $V$. 
We refer to these points as the singularity of the potential $V$ with respect to the momentum $p$, denoted as $q_0(p)$. 
While solving the scattering equation, we establish an integration path concerning momentum as follows. 
On this fixed integration path $C_p$, the potential function $V$ generated a path of singularity $C_{q_0}\left(C_p\right)$, which dependents on $C_p$.
If the integration path intersects with the path of $q_0$, there will be a discontinuity at the intersection point, which is not the desired outcome. 
Therefore, in the subsequent search of singularities, we designed a path $C_p$ on the complex momentum plane which avoid intersecting with the path of the singularities $C_{q_0}\left(C_p\right)$ generated by $V$, as shown in Fig.~\ref{fig:int_path}.

In Fig.~\ref{fig:int_path}, the solid line is the integrating path for searching the pole, the dashed lines are the path of $C_{q0}\left(C_p\right)$, which demonstrated that the poles of the potential $V_{\alpha\gamma}\left(p,q\right)$ with the expression Eq.~(\ref{equ:BSE_eq}), depending on a given complex $p$, and the isolated red point is the on-shell momentum of the energy at the pole position of $T-$matrix. 
The integration path is divided into 2 parts: the first part will go around the propagator pole; the second part will go to infinity and will take an angle such that it will not cross the potential-pole lines. 
Now based on this integration routine, we find the pole positions as shown in Table~1.
The integrating path for channel $\pi^0 P_c(4312)$ is on the second Riemann sheet, and the path for channel $\bar{D}^0\Lambda_{c1}^{+}$ is on the first Riemann sheet.
{\color{black} Here we should mention that in this numerical method, there are two Riemann sheets for each channel, named as first Riemann sheet and the second Riemann sheet, which are distinguished by the integral routine above or below the pole position of on-shell momenta of the channel at the pole energy.
Such pole position is defined by $E-\omega_{\gamma_B}(q)-\omega_{\gamma_M}(q)=0$, then the value of integration will change by these two choices of integral routines.
}
Thus, the poles correspond to the bound states of $\bar{D}^0\Lambda_{c1}^{+}$ while the resonances to $\pi^0 P_c(4312)$, the decay width is mainly for decaying to $\pi^0 P_c(4312)$ channel.
In the previous discussions, there is only one undetermined parameter, $\Lambda$, which indicates the high energy detail of the interaction, thus, we take several values to estimate the uncertainties of it.
As shown in Table~1, the changing cutoff does not affect the energy pole much.
It has been observed that the dominant contribution for potential arises from the small momenta, since the energy considered is rather close to the threshold. 
Thus, as shown in Eq.~(\ref{eq:cut}), we expect the form factor to approach a constant 1 as $p,q \to 0$.

\vskip 2mm \tabcolsep 4.5pt

\tl{1}\tabtitle{7.8}{1}{The pole position of $T-$matrix in the complex plain for different cutoffs.}

\centerline{\footnotesize
\begin{tabular}{cccccc}
\hline\hline\hline
& &
$M_{P_c}/\mev$
& & $\Gamma_{P_c}/2\mev$ \\
\hline
$\Lambda=0.8\gev$ &   & 4456.7428& & 10.7337 \\
$\Lambda=1.0\gev$ &   & 4456.7667& & 10.7293 \\
$\Lambda=1.2\gev$ &   & 4456.7861& & 10.7238\\
\hline\hline\hline
\end{tabular}
}

\vskip 2mm

\medskip

Based on this two-coupled-channel picture, we find the $P_c(4457)$ could be a bound state of $\bar{D}^0\Lambda^+_{c1}$, while it can decay to $\pi^0 P_c(4312)$.
{\color{black}In Ref.~\cite{Ling:2021lmq}, they also discuss the decay width of $P_c(4457)$ to $\pi^0 P_c(4312)$, however, in their model, it is $P$-wave decay since $P_c(4457)$ is a negative parity state.
Thus, the decay width is rather small in their work.}
The next step will be determining this outcome.
In the previous decay process, $\Lambda_b^0 \to J/\psi K^- p$, three $P_c$ states were found in the $J/\psi p$ invariant mass spectrum.
If $P_c(4457)$ can decay to $P_c(4312)\pi^0$, we want to propose a reaction, $\Lambda_b^0 \to J/\psi K^- p \pi^0$, where $J/\psi p \pi^0$ three-body is from $P_c(4457)$ while $J/\psi p$ two-body is from $P_c(4312)$.
However, the efficiency of $\pi^0$ is very low for the detector in LHCb, so we also propose another reaction, $\Lambda_b^0 \to J/\psi K_s p \pi^-$.
We hope the neutral partner of $P_c(4457)$ in the invariant mass spectrum of $J/\psi p \pi^-$, and correspondingly, $J/\psi p$ is still from $P_c(4312)$. 
However, from $\Lambda_b^0 \to J/\psi K^- p$, we only find several hundred events from $P_c(4457) K^-$, thus, we do not expect to find $\Lambda_b^0 \to J/\psi K_s p \pi^-$ in the current database from the Run-2 of LHCb facility.
When Run-3 data are available, we expect some signals in these processes.
Furthermore, $\Lambda_b^0 \to J/\psi p \pi^- \pi^+ K^-$ may be another choice, even though there are five final states, as all of these final states are charged particles, resulting in a high detection efficiency.

In this Letter, we propose a new mechanism to interpret $P_c(4457)$.
We investigate a two-coupled-channel system, including $\bar{D}^0\Lambda^+_{c1}$ and $\pi^0 P_c(4312)$.
We only considered the contribution from the off-diagonal elements of the potential matrix, due to the nearly on-shell exchange particle $\Sigma_c^+$. 
Through this interaction, we observed a singularity in the complex energy plane of this two-channel system, which is located on the first Riemann sheet of the $\bar{D}^0\Lambda^+_{c1}$ and on the second Riemann sheet of $\pi^0 P_c(4312)$, corresponding to a bound state of $\bar{D}^0\Lambda^+_{c1}$ and a resonance state of $\pi^0 P_c(4312)$. 
The position of this singularity is at $4457+i11 \mev$, and we also observe that the position of this state is not significantly related to the cutoff parameter. 
This pole position coincides exactly with $P_c(4457)$.
Therefore, we speculate that $P_c(4457)$ is the result of the interaction of these two coupled channels, and the particle has $J^p=1/2^+$. 
In contrast to the previous work, where the parity of $P_c(4457)$ is negative as for the bound state of $\bar{D}^*\Sigma_c$ under $S-$wave interaction, we predict that the parity of $P_c(4457)$, as of the $P_c(4312)\pi$-$\bar{D}\Lambda^+_{c1}$ $S-$wave interaction, is positive.
Therefore, determining the parity of this state in experiments can distinguish these two mechanisms.
Since our estimation was preliminary, further confirmation of the properties of $P_c(4457)$ requires additional experimental information and theoretical effort.
In addition, we have proposed three processes $\Lambda_b^0 \to J/\psi K_s p \pi^-$, $\Lambda_b^0 \to J/\psi K^- p \pi^0$ and $\Lambda_b^0 \to J/\psi K^- p \pi^+\pi^-$ for the future practice of facilities.

\textit{Acknowledgements.} 
The authors want to thank the useful discussions with Li-sheng Geng and Feng-kun Guo. 
This work is supported by the National Natural Science Foundation of China under Grant Nos. 12175239, 12135011, and 12221005,
and by the National Key Research and Development Program of China under Contracts 2020YFA0406400,
and by the Chinese Academy of Sciences under Grant No. YSBR-101, 
and by the Xiaomi Foundation / Xiaomi Young Talents Program.

\bibliographystyle{unsrturl}
\bibliography{ref}

\end{document}